\documentclass[
reprint, 
preprintnumbers,
notitlepage, 
aps,
prl,
superscriptaddress
]{revtex4-2}
\usepackage{amsmath}
\allowdisplaybreaks 
\usepackage{amssymb}
\usepackage{amsthm}
\usepackage{bibentry}
\usepackage{bm}
\usepackage{cancel}
\usepackage{dcolumn}
\usepackage{dsfont} 
\usepackage{graphicx}
\usepackage{hyperref}
\usepackage{lmodern}
\usepackage{mathrsfs}
\usepackage{physics}
\usepackage{simplewick}
\usepackage[colorinlistoftodos,prependcaption,textsize=small]{todonotes}
\usepackage[normalem]{ulem}
\usepackage{xcolor}

\usepackage{xcolor}
\definecolor{fg}{RGB}{34, 139, 34} 

\begin{document}


\title{Memory Function Representation for the Electrical Conductivity of Solids}

\author{Brett R. Green\; }
\affiliation{%
Department of Physics,
The Pennsylvania State University, Univeristy Park, PA 16802, USA}%
\author{Maria Troppenz}
\affiliation{Institut für Physik und Iris Adlershof, Humboldt-Universität zu Berlin,
Zum Grossen Windkanal 2, 12489, Berlin, Germany}
\author{Santiago Rigamonti}
\affiliation{Institut für Physik und Iris Adlershof, Humboldt-Universität zu Berlin,
Zum Grossen Windkanal 2, 12489, Berlin, Germany}
\author{Claudia Draxl}
\affiliation{Institut für Physik und Iris Adlershof, Humboldt-Universität zu Berlin,
Zum Grossen Windkanal 2, 12489, Berlin, Germany}
\author{Jorge O. Sofo\; }
\email{sofo@psu.edu}
 \homepage{http://sites.psu.edu/sofo}
\affiliation{%
Department of Physics,
The Pennsylvania State University, Univeristy Park, PA 16802, USA}%
\affiliation{%
Department of Materials Science and Engineering, and Materials Research Institute,
The Pennsylvania State University, Univeristy Park, PA 16802, USA}%

\date{\today}

\begin{abstract}
We derive a formula for the electrical conductivity of solids that includes relaxation, dissipation, and quantum coherence. 
The derivation is based on the Kubo formula, with a Mori memory function approach to include dissipation effects at all orders in the relaxation interaction.
It offers a practical method to evaluate the conductivity with electronic-structure codes and avoids the complications and limitations of the Kubo formula in the thermodynamic limit. 
The derivation of our formula provides a method applicable to other transport coefficients and correlation functions.
\end{abstract}

\maketitle

Most \textit{ab initio} evaluations of the conductivity are based on the Boltzmann equation \cite{scheidemantelTransportCoefficientsFirstprinciples2003,madsenBoltzTraP2ProgramInterpolating2018,ponceFirstprinciplesCalculationsCharge2020}. 
Although originally derived for a classical gas, it can be obtained from quantum dynamics as an equation for the diagonal elements of the density matrix \cite[Chpts. 6 and 7]{kadanoffQuantumStatisticalMechanics1989}. 
Neglecting quantum coherence, represented by the non-diagonal elements of the density matrix, is deemed as a good approximation for weak scattering. 
In this Letter, we provide a formula for the electrical conductivity of solids that is valid for all scattering strength and includes quantum coherence effects. Our formula recovers the Boltzmann result for weak scattering and provides a quantitative assessment of its range of validity in terms of scattering strength and band structure of the material.  

Our derivation starts from the Kubo formula \cite[Eq. (5.11)]{kuboStatisticalMechanicalTheoryIrreversible1957a} that provides a rigorous starting point for the evaluation of linear transport coefficients. 
However, this formula is only applicable to systems that display a linear response to the perturbation considered.
In supercell-based representations of disorder, it gives exact results only in the thermodynamic limit, {\it i.e.}, when macroscopic systems are used for its evaluation. 
However, this limit is impossible to achieve with electronic-structure codes, and approximation methods for the relaxation processes are needed. 
To overcome this limitation, we perform a Hilbert-space projection to obtain a generalized Langevin equation \cite[]{moriTransportCollectiveMotion1965}. 
This equation describes the evolution of density excitations of the electron system when subject to a periodic potential with static or dynamical perturbations which break conservation of momentum and energy, such as disorder or phonons. 
We keep excitations of a given wave vector $\bm{q}$ as our active subspace and project those of different wave vectors $\bm{q}^{\prime}\ne\bm{q}$ into a bath described by a memory function.
We use conservation laws to define slower variables within the active subspace. This facilitates the approximation of the dissipation process into a single memory function \cite[]{gotzeMobilityQuantumParticle1981}. 
All this will be derived in more detail below, however the importance of the main result can be appreciated at this point.
The resulting expression for the optical conductivity in the presence of the perturbations is
\begin{align}
  \label{eq:optical-conductivity}
  \sigma_{\mu,\nu}(\omega)=\sigma^{(0)}_{\mu,\nu}(\omega-M^{\prime}(\omega)-iM^{\prime\prime}(\omega))\;,
\end{align}
where $\sigma^{(0)}(z)$, with $z$ in the complex plane, is the optical conductivity evaluated for the periodic system and $M^{\prime}(\omega)$ ($M^{\prime\prime}(\omega)$) is the real (imaginary) part of the memory function on the real axis in the $\bm{q}\to 0$ limit. 
From this expression, we can obtain the following formula to evaluate the electrical conductivity in the dc ($\omega \rightarrow 0$) limit,
\begin{widetext}
\begin{align}
  \label{eq:sigma-dc}
  \sigma_{\mu,\nu}^{(\text{dc})}=\frac{1}{M^{\prime\prime}}\sum_{n,\bm{k}}\qty(\frac{\partial f}{\partial \varepsilon})_{\varepsilon_{n,\bm{k}}} 
  \frac{\partial \varepsilon_{n,\bm{k}}}{\partial k_{\mu}}
  \frac{\partial \varepsilon_{n,\bm{k}}}{\partial k_{\nu}}-
 &\sum_{n,\bm{k}\alpha}f(\varepsilon_{n,\bm{k}})\epsilon_{\mu,\nu,\alpha}\Omega_{n;\alpha}(\bm{k})+
 \nonumber\\
 & M^{\prime\prime}\sum_{\substack{n^\prime,n,\bm{k}\\n\ne n^{\prime}}}
  X_{n,n^{\prime};\mu}(\bm{k})X_{n^{\prime},n;\nu}(\bm{k})
  \frac{\qty[f(\varepsilon_{n,\bm{k}})-f(\varepsilon_{n^{\prime},\bm{k}})]
  \qty(\varepsilon_{n,\bm{k}}-\varepsilon_{n^{\prime},\bm{k}})}
  {M^{\prime\prime 2}+\qty(\varepsilon_{n,\bm{k}}-\varepsilon_{n^{\prime},\bm{k}})^2}\;,
\end{align}
\end{widetext}
where $M^{\prime\prime}:=M^{\prime\prime}(0)$, $f(\varepsilon)$ is the Fermi occupation function, $\varepsilon_{n,\bm{k}}$ represents the band energies of the periodic system, $X_{n^{\prime},n;\alpha}(\bm{k})
=i\bra{u_{n^{\prime},\bm{k}}}\ket{\partial u_{n,\bm{k}}/\partial k_{\alpha}}$
is the interband part of the position matrix element in the Bloch basis \cite{karplusHallEffectFerromagnetics1954}, $\bm{\Omega}_{n}(\bm{k})=\operatorname{curl}_{\bm{k}} \bm{X}_{n,n}(\bm{k})$
is the the Berry curvature of band $n$, and $\epsilon_{\mu,\nu,\alpha}$ is the Levi-Civita symbol.

Before we embark on its derivation, there are important consequences of this formula worth mentioning. 
The first term in Eq.~(\ref{eq:sigma-dc}) is the result obtained from the Boltzmann transport equation, see {\it e.g.}, Refs. \cite{scheidemantelTransportCoefficientsFirstprinciples2003,madsenBoltzTraP2ProgramInterpolating2018,ponceFirstprinciplesCalculationsCharge2020}.
For a parabolic band with an effective mass $m^{*}$, the first term becomes the Drude result $\sigma_{x,x}^{(\text{dc})}=n\tau/m^*$ if we associate $M^{\prime\prime}=1/\tau$. 
If the Fermi level intersects bands that are non-parabolic, the derivative of the dispersion relation accounts for the non-parabolicity. 
The first term in Eq.~(\ref{eq:sigma-dc}) provides the contribution to the conductivity due to intraband transitions from each of those bands. 


Importantly, the third term in Eq.~(\ref{eq:sigma-dc}) shows that in a multiband system, there is an additional interband contribution. 
Low-energy optical transitions in the system contribute to the dc conductivity due to the broadening produced by the scattering. 
This broadening is represented by the memory function.
Non-diagonal terms of the density matrix are important for the correct description of the non-equilibrium transport process. 
In derivations of the Boltzmann transport equation from the exact dynamics \cite{kadanoffQuantumStatisticalMechanics1989}, this interband coherence was neglected because the focus was on the free electron gas or similarly simple systems. 
The third term in Eq.~(\ref{eq:sigma-dc}) is a means to explore the importance of interband coherence on the transport process and to corroborate or disprove the modern common assumption \cite[paragraph after Eq.(20)]{ponceFirstprinciplesCalculationsCharge2020} that they are not important. 
Indeed, this term becomes important when the system has one or more bands at an energy of the order of $M^{\prime\prime}$ from the Fermi energy. 
This sets the limits of validity of Boltzmann's approach and makes evident that this term should be included in any general tool for the evaluation of electrical conductivity from electronic-structure calculations.
Even if the scattering is weak and well captured by a relaxation time, the effects described by the third term in Eq.~(\ref{eq:sigma-dc}) can be important due to the presence of more than one band close to the Fermi level. 
It should be taken into account for computing the conductivity from first principles.
We consider this third term and the physical origin of broadening in $M$ to be the two most important results presented here.

There are two possible scenarios where this third term of Eq.~(\ref{eq:sigma-dc}) becomes important. 
One situation is the case of strong scattering, where $M^{\prime\prime}$ becomes large and more bands of the pristine system are involved in the scattering. 
The other situation concerns calculations with large supercells.
To include the effect of disorder, we create supercells of the original material and include disorder in the supercell.
Large supercells lead to a multiple folding of the original Brillouin zone, creating more bands at smaller energy separation. Even in the case of weak scattering, the third term will become more and more important as we increase the size of the supercell. 
The memory function is dependent on supercell size and has to be recalculated for each supercell.

The second term represents the anomalous conductivity in system with a nonzero Berry curvature; it is zero for the diagonal elements of the conductivity tensor. 
It is interesting to note that the frequency independence of the anomalous conductivity, second term in Eq.~(\ref{eq:sigma-dc}), shows that to this level of approximation, this term is not affected by scattering. 
Whether this is a consequence of topological protection remains an open question. 

As a whole, Eq.~(\ref{eq:sigma-dc}) provides a formula for the electrical conductivity that is valid for any scattering strength as long as $M^{\prime\prime}$ is calculated consistently with the theory.
In the limit of strong scattering, when $M^{\prime\prime}(0)\to\infty$, the conductivity is zero, indicating the metal-insulator transition. 
We will explore this limit in a future publication where we present a self-consistent method to determine the memory function from a disorder potential. 
This divergence of the memory function can be produced by a finite amount of disorder and captures the Anderson transition \cite{gotzeMobilityQuantumParticle1981}. 


We start our derivation from the Kubo formula for the electrical conductivity \cite[Eq. (5.11)]{kuboStatisticalMechanicalTheoryIrreversible1957a} given by 
\begin{subequations}
  \label{eq:sigma-def}
  \begin{align}
    \label{eq:lt-cond}
    \sigma_{\alpha,\beta}(\bm{q},z)
    &=-i \int_{0}^{\infty}dt \ \! e^{izt}\sigma_{\alpha,\beta}(\bm{q},t)\\
    \label{eq:int-prod-curr}
    \sigma_{\alpha,\beta}(\bm{q},t)&=i\left(j_{\alpha}(\bm{q}),j_{\beta}(\bm{q},t)\right)
    \nonumber\\
    &=i\int_{0}^{\beta}d\lambda \expval{e^{\lambda H}j_{\alpha}(-\bm{q})e^{-\lambda H}j_{\beta}(\bm{q},t)}
  \end{align}
\end{subequations}
where $\beta$ is the inverse temperature, and $\expval{\ldots}$ is the grand-canonical equilibrium average at that temperature. 
Eq.~(\ref{eq:int-prod-curr}) introduces an internal product between operators that we will use to define projectors. 
It has all required properties of an internal product in operator space, appears naturally in the derivation of this formula, and minimizes the quantum noise in the averaged Langevin equation that we will use in our derivation \cite[p. 155]{zwanzigNonequilibriumStatisticalMechanics2001}.
The operators involved are components of the particle-current operator, $\bm{j}(\bm{q},t)$ at time $t$. We omit the time variable to indicate the operator at $t=0$. 
The time evolution is driven by the Hamiltonian of the system for zero external electric field, $H$.
This expression manifests the conceptual definition of the time-dependent conductivity as the overlap between the current at time $t$ with the current at time $0$.

The Kubo formula gives the exact linear response of a system, {\em as long as there is a linear regime} and the Hamiltonian for the system is known and solvable. 
These are not trivial conditions to meet. 
For example, a supercell calculation of a disordered alloy with random composition inside the supercell is a periodic system. 
For any finite size of the supercell, this system is going to be a perfect conductor, {\it i.e.}, it will not reach a steady-state current when subject to a homogeneous and constant electric field. 
This is manifested by a divergent value of the conductivity at zero frequency which is certainly not what we expect of a {\em linear} response.
Only in the thermodynamic limit, when the system has a macroscopic number of degrees of freedom, the linear-response calculation is valid. 
This fact was earlier recognized by Greenwood when after \cite[Eq. (29)]{greenwoodBoltzmannEquationTheory1958} points out that "...a large number of energy levels are included" into the "quasi-$\delta$-function" of the now known as the Kubo-Greenwood formula.
If we try to approach this limit through supercells of increasing size, at each step the system is periodic and does not have a finite linear response; the current diverges as we wait infinite time to reach equilibrium. 
This is a case where the system does not have a linear regime.

These problems can be shown explicitly by evaluating the conductivity, Eq.~(\ref{eq:sigma-def}), for a periodic supercell. 
The Hamiltonian is diagonal in Bloch states,  $\ket{n,\bm{k}}$, created by the operator $c^\dagger_{n\bm{k}}$ and can be written as 
$H_0=\sum_{n\bm{k}} \varepsilon_{n,\bm{k}}c^\dagger_{n\bm{k}}c^{ }_{n\bm{k}}$,
where $\varepsilon_{n,\bm{k}}$ is the single particle energy of crystal momentum $\bm{k}$ and band $n$ measured from the chemical potential $\mu$. 
The diagonal conductivity for a uniform electric field, $\bm{q}\to 0$, is
\begin{align}
  \label{eq:sigma0}
  &\sigma^{(0)}_{\alpha,\alpha}(\omega+i0^{+})=-\frac{i}{\omega+i0^{+}}\sum_{n\bm{k}}\qty(\frac{\partial f}{\partial \varepsilon})_{\varepsilon_{n,\bm{k}}} 
  \qty(\frac{\partial \varepsilon_{n,\bm{k}}}{\partial k_{\alpha}})^{2}
  \nonumber\\
  &\phantom{=}+ i\omega\sum_{\substack{n^\prime,n,\bm{k}\\n\ne n^{\prime}}} \qty|X_{n,n^{\prime};\alpha}(\bm{k})|^{2}\frac{f(\varepsilon_{n,\bm{k}})-f(\varepsilon_{n^{\prime},\bm{k}})}{\omega+i0^{+}+\varepsilon_{n,\bm{k}}-\varepsilon_{n^{\prime},\bm{k}}}\;.
\end{align}
From this expression, we can see that the response of any periodic supercell to a constant and uniform electric field is divergent due to the $\delta(\omega)$ appearing in the first term. 
As the size of the supercell increases, weight from this pole at zero frequency is transferred to interband transitions at finite frequencies represented by the second term.
These frequencies become smaller and smaller as the size of the supercell increases.
For weak scattering, these transitions form the Drude peak in the thermodynamic limit. However, very large supercells are required to get the correct value of the dc conductivity if convergence can be reached at all.
This problem is circumvented by considering an evaluation of the Kubo formula that takes into account scattering to all orders as we describe in what follows.

The derivation of our main result, Eq.~(\ref{eq:optical-conductivity}), starts by recognizing that the elementary excitation in a system of Fermions are electron-hole creations represented by the operator $\xi_{n^\prime,n,\bm{k}}(\bm{q})=c^\dagger_{n^\prime,\bm{k}-\bm{q}} c^{ }_{n,\bm{k}}$.
These elementary excitations of momentum $\bm{q}$ form an invariant subspace of the Hamiltonian $H_{0}$ and consequently have an infinite lifetime due to momentum conservation in a system without scattering.
However, if the particles scatter due to impurities, phonons, or any other effect that breaks periodicity, the total Hamiltonian will mix these excitations of momentum $\bm{q}$ with others of different momentum, $\bm{q}^{\prime}\ne\bm{q}$.
Hence, excitations with a given momentum $\bm{q}$ acquire a finite lifetime by mixing with other excitations of different momentum. 
The effect of this interaction on the dynamics of the excitations with momentum $\bm{q}$ will be treated with the memory function formalism \cite{moriTransportCollectiveMotion1965}. 

Since the elementary excitations are not orthogonal with respect to the inner product defined in Eq.~(\ref{eq:int-prod-curr}), it is convenient to define dynamical variables for the subspace of excitations with wave vector $\bm{q}$ through linear combinations of the form 
\begin{align}
  A_{\alpha}(\bm{q})
  =\sum_{n,n^{\prime},\bm{k}}a^{(\alpha)}_{n,n^{\prime},\bm{k}}(\bm{q})\xi_{n,n^{\prime},\bm{k}}(\bm{q})
  \;.
\end{align}
The coefficients should be chosen to enforce orthogonality.
Note that both density and current fluctuations of wave vector $\bm{q}$ can be written in this form.
The time evolution of these variables, $A_{\beta}(\bm{q},t)=e^{i{\cal L}t}A_{\beta}(\bm{q})$, is described with a Liouville operator ${\cal L}\bullet=[H,\bullet]$. 
In order to separate the time evolution within the $\bm{q}$ subspace from that of excitations with other wave vectors, we define the projector into the $\bm{q}$ subspace ${\cal P}(\bm{q})\bullet=\sum_{\alpha} A_\alpha(\bm{q}) \qty(A_\alpha(\bm{q}),\bullet)$, as well as its complement ${\cal Q}(\bm{q})=1-{\cal P}(\bm{q})$.

Instead of the time evolution of the operators, it is more convenient to evaluate the time evolution of correlation functions of the form
\begin{align}
  \label{eq:corr-def}
  \phi_{\alpha,\beta}(\bm{q},t)
  &=\left(A_{\alpha}(\bm{q}),A_{\beta}(\bm{q},t)\right)\;,
\end{align}
because the random force does not appear in their projected time evolution  \cite[Eq. (3.11)]{moriTransportCollectiveMotion1965}; the random force is orthogonal to the variable at $t=0$, $A_{\alpha}(\bm{q})$.
In particular, the conductivity, Eq.~(\ref{eq:sigma-def}), is a correlation function of this general form.
As can be seen in Eq.~(\ref{eq:corr-def}), the relaxation function measures the amplitude of a given excitation after a certain time $t$ has passed during the evolution of the system, just as the conductivity of Eq.~(\ref{eq:lt-cond}) measures the relaxation of current excitations.

Following Mori \cite{moriTransportCollectiveMotion1965}, we write the equation of motion of these correlation functions as a generalized Langevin equation, which after a Laplace transform reads
\begin{align}
  \label{eq:eom-Phi}
  \sum_{\mu}\Phi_{\alpha,\mu}(\bm{q},z)
  \qty[z\delta_{\mu,\beta}+\Omega_{\mu,\beta}(\bm{q})-M_{\mu,\beta}(\bm{q},z)]=\delta_{\alpha,\beta},
\end{align}
where $\Phi_{\alpha,\mu}(\bm{q},z)=-i\int_{0}^{\infty}dte^{izt}\phi_{\alpha,\mu}(\bm{q},t)$ is the Laplace transform of the relaxation function and $\Omega_{\mu,\beta}(\bm{q})=\qty(A_{\mu}(\bm{q}),{\cal L}A_{\beta}(\bm{q}))$ 
is a static correlation representing the frequency of oscillation of the collective variables without dissipation. Dissipation is included through the definition of the memory function, given by
\begin{align}
  \label{eq:mem-def}
  M_{\mu,\beta}(\bm{q},z)=\qty(Q(\bm{q})LA_{\mu}(\bm{q}),
  R_{Q(\bm{q})}(z)Q(\bm{q})LA_{\beta}(\bm{q}))\;,
\end{align} 
with $R_{Q(\bm{q})}(z)=\qty[z+Q(\bm{q})L]^{-1}$.
The memory function is the correlation of the random force ${\cal Q}{\cal L}A_{\mu}(\bm{q})$ evolving in time with the propagator $R_{Q(\bm{q})}(z)$ in the subspace of the bath, which corresponds to excitations of wave vector $\bm{q}^{\prime}\ne\bm{q}$.

Although the equation of motion, Eq.~(\ref{eq:eom-Phi}), is very appealing in structure because of the similitude with the Langevin equation, it is still exact and hence only solvable in simple cases.
However, we can use conservation laws, such as particle conservation, to detect slow variables and find simplifying approximations. 
Conserved quantities are slow because their change has to be associated with the transport of that quantity through a current \cite[p. 11]{forsterHydrodynamicFluctuationsBroken1975}. 
Particle conservation, for example, is represented by the continuity equation that shows that the time evolution of the density is equal to the divergence of the current. 
In terms of the spatially Fourier transformed operators, the continuity equation is ${\cal L}\rho(\bm{q})=-\bm{q}\cdot\bm{j}(\bm{q})$,
where $\rho(\bm{q})=\sum_{n^{\prime},n,\bm{k}}M_{n^{\prime},n,\bm{k}}(\bm{q})\xi_{n^{\prime},n,\bm{k}}(\bm{q})$
creates a density excitation of momentum $\bm{q}$, and 
$M_{n^{\prime},n,\bm{k}}(\bm{q})=\bra{u_{n^{\prime},\bm{k}-\bm{q}}}\ket{u_{n,\bm{k}}}$ is the overlap of the periodic parts of two Bloch states.
This consideration suggests that we can choose the normalized particle-density fluctuation as our first collective variable
\begin{align}
  \label{eq:A0}
  A_{0}(\bm{q})=\frac{1}{\sqrt{\qty(\rho(\bm{q}),\rho(\bm{q}))}}\rho(\bm{q})\;.
\end{align}
Naturally, since for any operator the inner product obeys $\qty(A,{\cal L}A)=0$, we can choose our next variable as
\begin{align}
  A_{1}(\bm{q})=\frac{1}{\sqrt{\qty({\cal L}\rho(\bm{q}),{\cal L}\rho(\bm{q}))}}{\cal L}\rho(\bm{q})\;,
\end{align} 
which is proportional to the longitudinal current and orthogonal to $A_{0}(\bm{q})$. 
The most consequential result of these choices is that, due to the continuity equation, the action of ${\cal L}$ on the charge fluctuations of momentum $\bm{q}$ does not mix them with fluctuations of other wave vectors, so that the action of the projector ${\cal Q}(\bm{q})$ is zero.
This means ${\cal Q}(\bm{q}){\cal L}\rho(\bm{q})=0$, and all the rows and columns of the memory function related to the charge are zero, {\it i.e.}, $M_{\alpha,0}(\bm{q},z)=M_{0,\beta}(\bm{q},z)=0$ for all $\alpha$ and $\beta$.
As a consequence of particle conservation, the diagonal element of the memory function corresponding to the longitudinal current, $M_{1,1}(\bm{q},z)$, dominates the dynamics.  
Following an approximation devised by G\"otze \cite{gotzeMobilityQuantumParticle1981}, we write the memory-function matrix as a diagonal matrix with $0$ in the first diagonal term and $M(\bm{q},z):=M_{1,1}(\bm{q},z)$ in the rest of the diagonal. Additionally, we evaluate all static correlation functions in the reference system without scattering. 
With these two approximations and after a little algebra, the equation of motion, Eq.~(\ref{eq:eom-Phi}), can be solved for the normalized density correlation function
\begin{align}
  \label{eq:Phi00-mem}
  \Phi_{0,0}(\bm{q},z)=\frac{\Phi_{0,0}^{(0)}(\bm{q},z-M(\bm{q},z))}{1+M(\bm{q},z)\Phi_{0,0}^{(0)}(\bm{q},z-M(\bm{q},z))}\;.
\end{align} 
Here $\Phi_{0,0}^{(0)}(\bm{q},z)$ is the solution to Eq.~(\ref{eq:eom-Phi}) for zero memory function. 
It is the normalized density correlation function for the system without scattering, {\it i.e.}, the periodic supercell.

With Eq.~(\ref{eq:Phi00-mem}) we can calculate the density response function if we know the memory function. 
The definition of the memory function, Eq.~(\ref{eq:mem-def}), provides a starting point for different approximations. For weak scattering, it can be evaluated using perturbation theory. However, it can also be used to obtain quite general expressions that are valid for all strengths of the scattering, such as the  mode-coupling theory \cite{gotzeMobilityQuantumParticle1981}. 
These methods and approximations will be discussed in a future publication. 
At this point, we want to focus our attention on the expression obtained for the correlation functions, specifically for the conductivity. 

The continuity equation provides the link between the density response function and the longitudinal conductivity,
\begin{align}
  \label{eq:sigma-pol}
  \sigma(\bm{q},z)=i\frac{zg(\bm{q})}{q^2}\qty[z\Phi_{0,0}(\bm{q},z)-1]=-i\frac{z}{q^2}\chi(\bm{q},z)\;,
\end{align}
where $g(\bm{q})=\qty(\rho(\bm{q}),\rho(\bm{q}))$ and in the last identity we have introduced the density correlation function $\chi(\bm{q},z)$ \cite[]{forsterHydrodynamicFluctuationsBroken1975,kadanoffHydrodynamicEquationsCorrelation1963}. 
The preceding expressions are derived by operator identities and are of general applicability to any system. 
In accordance with the approximation made for the static correlation functions, we take $g(\bm{q})=g^{(0)}(\bm{q})$ that is evaluated with the periodic Hamiltonian.
In the limit of small $q$, the density correlation function takes the form $\chi(\bm{q},z)\approx \bm{q}\cdot\bm{\alpha}(z)\cdot\bm{q}$ where $\bm{\alpha}(z)$ is the dipole-dipole response or polarizability \cite[Eq.~(4)]{gotzeTheoryConductivityFermion1979}. 

Now, if we introduce the expression of the density relaxation function, Eq.~(\ref{eq:Phi00-mem}), into the expression for the conductivity, Eq.~(\ref{eq:sigma-pol}), and take the homogeneous limit $q \rightarrow 0$, we find the optical conductivity given by
\begin{align}
  \label{eq:final-result}
  \sigma_{\mu,\nu}(z)&=-i\qty[z-M(z)]\alpha_{\mu,\nu}^{(0)}(z-M(z))
  \nonumber\\
  &=\sigma_{\mu,\nu}^{(0)}(z-M(z))\;,
\end{align}
where $M(z)=\lim_{q\to 0}M(\bm{q},z)$.
This is our main result, introduced previously in Eq.~(\ref{eq:optical-conductivity}).
This expression is independent of the method used to evaluate the memory function and introduces relaxation effects into the calculation of the optical response of materials \cite{ambrosch-draxlLinearOpticalProperties2006}. 

In addition to describing the effects of scattering produced by disorder or phonons on the optical response, our method allows for the correct dc limit. 
The dc conductivity is given by 
\begin{align}
  \sigma_{\mu,\nu}^{(\text{dc})}=\lim_{\omega\to 0}\lim_{\eta\to 0}\Re \sigma_{\mu,\nu}(\omega+i\eta)\;.
\end{align}
In this limit, the memory function is purely imaginary and given by $iM^{\prime\prime}$. 
Consequently, the dc conductivity is given by Eq.~(\ref{eq:sigma-dc}).
Note that the real part of the memory function at zero frequency is zero because it is the Fourier transform of a real function \cite{forsterHydrodynamicFluctuationsBroken1975}.

Note that, in the lowest order correction with respect to the scattering strength, our result coincides with the perturbative treatment of Kohn and Luttinger \cite[]{kohnQuantumTheoryElectrical1957}.
Also, exact sum rules obeyed by this formula show that for weak scattering, the inertial mass of the carriers is the effective mass (curvature of the band dispersion relation) while for strong scattering, where the electrons become more and more localized, it is the free electron mass.

It is clear from our result that the evaluation of the Kubo formula for supercells does not capture the relaxation processes properly. 
These processes only appear in the thermodynamic limit and, as our derivation shows, are represented by the memory function. 
Note that $M^{\prime\prime}$ in Eq.~(\ref{eq:sigma-dc}) adopts the same role as the small imaginary part used in the derivation of the Kubo formula for supercells \cite[Eq. (2.21)]{kuboStatisticalMechanicalTheoryIrreversible1957a}, also present in Eq.~(\ref{eq:sigma0}) when we set $z\to i \eta$. 
However, the latter has to be set to $0^{+}$ while the former has a finite value calculated from the scattering process. 
This observation explains the dc conductivity's pathological dependence on the small imaginary part when the Kubo formula is applied to supercells without further consideration of the thermodynamic limit \cite{calderinKuboGreenwoodElectrical2017,sangalliOpticalPropertiesPeriodic2017a}. 
The small imaginary part has to be kept finite as a poor approximation of the relaxation effects. 
This small imaginary part dominates the result of the calculation, invalidating all the effort put into the evaluation of the Kubo formula from first-principles calculations.

In summary, we have derived a formula, given in Eq.~(\ref{eq:final-result}), that allows the evaluation of the conductivity from results of electronic-structure codes and a memory function that incorporates the dissipation effects. 
These dissipation effects can be modeled at various levels of sophistication: from a simple constant relaxation time approximation to self-consistent evaluations capable of describing localization, phonon interactions, or other possible dissipation mechanisms.
The formula obtained generalizes the expression for a single band obtained in several treatments of the memory function approach \cite{gotzeHomogeneousDynamicalConductivity1972,defilippisAlternativeRepresentationKubo2014,allenElectronSelfenergyGeneralized2015} where the conductivity for the system without scattering is given only by the intraband contribution, the first term in Eq.~(\ref{eq:sigma-dc}). 
We are currently working on the consequences of this approach for the Seebeck coefficient and the electrical contribution to the thermal conductivity.

\begin{acknowledgments}
  B.G. acknowledges support and training provided by the Computational Materials Education and Training (CoMET) NSF Research Traineeship (Grant No. DGE-1449785) and support from the NSF Graduate Research Fellowship Program (Grant No. DGE1255832). M.T. acknowledges funding from the Elsa-Neumann Stiftung Berlin.
\end{acknowledgments}

\bibliography{PRL-mem-func-elec-cond-sol}

\end{document}